# The Potential of the Human Connectome as a Biomarker of Brain Disease

Marcus Kaiser

School of Computing Science, Newcastle University, UK;
Institute of Neuroscience, Newcastle University, UK; and
Department of Brain and Cognitive Sciences, Seoul National University, Korea

**Correspondence:**

Dr Marcus Kaiser
Newcastle University
School of Computing Science
Newcastle upon Tyne, NE1 7RU, UK
m.kaiser@ncl.ac.uk

*Abstract.* *The human connectome at the level of fiber tracts between brain regions has been shown to differ in patients with brain disorders compared to healthy control groups. Nonetheless, there is a potentially large number of different network organizations for individual patients that could lead to cognitive deficits prohibiting correct diagnosis. Therefore changes that can distinguish groups might not be sufficient to diagnose the disease that an individual patient suffers from and to indicate the best treatment option for that patient. We describe the challenges introduced by the large variability of connectomes within healthy subjects and patients and outline three common strategies to use connectomes as biomarkers of brain diseases. Finally, we propose a fourth option in using models of simulated brain activity (the dynamic connectome) based on structural connectivity rather than the structure (connectome) itself as a biomarker of disease. Dynamic connectomes, in addition to currently used structural, functional, or effective connectivity, could be an important future biomarker for clinical applications.*

**Keywords:** brain connectivity, network disease, brain disorders, classification, diagnosis

The study of how different components of the brain, may they be neurons or brain regions, are connected has become an emerging field within the neurosciences (Bullmore and Sporns, 2009; Kaiser, 2011; Sporns et al., 2004). The analysis of physical connections within neural systems gained momentum around 20 years ago with the availability of information on the nematode *Caenorhabditis elegans'* nervous system (Achacoso and Yamamoto, 1992; White et al., 1986) and the rhesus monkey's visual system of cortico-cortical connections (Felleman and van Essen, 1991; Young, 1992). Now called connectomics, the field aims to discover the structure of brain networks, representing physical connections such as axons or fiber tracts. As a next milestone, the first data sets of the Human Connectome Project are being released. What will the next 20 years bring? Like for genomics, the hopes are that features of the connectome of a patient can be a biomarker for diseases and an indicator for therapeutic interventions. Identifying biomarkers for diseases based on large-scale genome studies has been challenging. Is the link between connectivity and brain disease also over-weighted? What could a structural connectome in principle tell us about the brain organization in health and disease?

In analogy to genetics, we may distinguish a genotype and a phenotype of brain organization. The genotype is given by the structural connectivity either observed at the level of individual synapses (microconnectome) or at the level of fiber tracts between brain regions (macroconnectome) (DeFelipe, 2010) and we will refer to this as connectome. As for every novel field, the underlying techniques are still under development (Jbabdi and Johansen-Berg, 2011). Diffusion tensor and diffusion spectrum imaging can give us information on potential structural connections of the macroconnectome. The phenotype represents activity, as seen in fMRI or EEG, or behavior, as for cognitive clinical scores. We refer to these patterns as consequences on dynamics or behavior due to changed brain connectivity.

The problem of diagnosing a disease, as in genetics, is due to the fact that several mutations of the genotype might result in the same phenotype (disease). Observing brain connectivity, there might be several combinations of changes in fiber tracts leading to hallucinations or seizures, for example. Also, the same connectome organization might lead to different dynamics for changes that affect the internal anatomy and activity of network nodes but not the nodes' topology (Figure 1). The idea that many pathways can lead to similar behavior is linked to the concept of degeneracy (Price and Friston, 2002; Tononi et al., 1999), 'the ability of elements that are structurally different to perform the same function or yield the same output.' If the output (phenotype) is cognitive deficits in patients, the number of connectome (genotype) patterns that lead to such behavior can be seen as the degeneracy of a brain disease. Also, a higher degeneracy, meaning that more connectome patterns are linked to a disease, might result in a higher incidence in a population. A related observation has been made in the field of genetics when linking genetic changes to diseases: multiple genotypes might lead to the same phenotype (heterogeneity) (Addington and Rapoport, 2012). Therefore, detecting one connectivity pattern linked to a disease might only relate to a fraction of all patients. Moreover, many connectome changes will be neutral in that they do not lead to a brain disorder; thus variability in the healthy population is expected to be large as well. As for genetics, connectomics is currently moving to large-scale studies, e.g. the 1,200-subject Human Connectome Project or the 1,000-subject Functional Connectomes study, to address this underlying variability.

Another problem besides large connectome variability ('noise') is that cognitive deficits might arise from small changes ('signal'). Development can be seen as a system of nonlinear dynamics (Turing, 1952). It has become clear that genetic encoding (Kendler et al., 2011) and self-organization shape the formation of neural systems in health and disease. For self-organization, the interaction with the environment (external factors) or physical constraints (internal factors) can influence the establishment and survival of axonal connections. Consequently, small changes during development might lead to a different connectome and as a result to a different resulting consequence for cognition and behavior of human subjects. As the dynamics in the brain are also non-linear, a small change in structural connectivity might be sufficient to lead to changes in cognition and behavior. Relatively small changes in connectivity might be sufficient to lead to a brain disorder. Therefore, some connectivity patterns seen in patients might be quite close to the organization of healthy subjects.

Let us look at some cases of how brain diseases could be linked to brain connectivity. Also, let us only use two cases of how a network structure (edge or node) in a patient could differ from that of a control group: a significant increase or a significant decrease of a network measure. We will only look at a single measure here, say number of streamlines for edges and total strength of its connections for nodes, but our general observations also hold for a combination of network measures (Costa et al., 2007; Kaiser et al., 2009; van den Heuvel et al., 2012).

First, a disease might affect a single brain region which could have an effect on brain dynamics by changing its own activity pattern, the pattern of directly connected neighbors of the region, and, indirectly, the activity in the rest of the brain mediated by intermediate brain regions. As a simplification, let us assume that only structural connections from that brain region will be altered. As each brain region (for a parcellation in humans of 110 cortical and subcortical regions including both hemispheres) is connected to around 10 other brain regions, there are $2^{10} = 1,024$ possible changes assuming that each connection could either be significantly increased or decreased in a patient. Thus even at the local scale, only affecting a single brain region, many variations of a disease are possible.

Second, a disease could affect a set of network nodes. For example, regions of the neocortex mature at different times during development: medial regions before lateral regions and posterior before anterior regions. A change in the maturation of the frontal lobe could affect multiple regions at the same time and might affect a whole network module (Nisbach and Kaiser, 2007). Say that 10 regions show a different internal structure that also manifests itself in altered fiber tracts between them and other brain regions. Therefore, assuming 10 fiber tracts per brain region, or $10^2 = 100$ fiber tracts for all 10 affected regions, show changes leads to $2^{100} = 1.3 * 10^{30}$ variants. Let us look at a simpler model where an increase (or a reduction) in at least 10 of those 100 fiber tracts is sufficient to lead to the behavioral features of a disease. There are $\binom{100}{10} = 1.7 * 10^{13}$ ways to choose 10 out of 100 connections. Given that 10 is the lower bound for disease onset, choosing 11, 12, 13, etc. connections leads to even more variations at this regional level.

Third, a disorder could lead to changes of a set of edges at the global level as a result of widely distributed changes. If there are 500 bidirectional connections (fiber tracts) between our 110 brain regions, there are $2^{500} = 3.3 * 10^{150}$ possible changes compared to a benchmark brain based on a population of healthy subjects.

We know that there is huge variability not only in the surface shape of human brains but also in its related connectivity pattern (Hilgetag and Barbas, 2006; Van Essen, 1997). Clearly, only a small fraction of connectome patterns is linked to a brain disorder. Even if we assume that there are thousands of subtypes of brain disorders, e.g. different kinds of epilepsy, and that many diseases change synaptic efficacy without changing structural connectivity, there might still be billions of connectome changes that could lead to the clinical patterns observed in patients with one type of a disease. Clearly, no two patients are the same (neither are no two control subjects).

If there is a multitude of ways how connectome changes could lead to a disease, how can we use brain connectivity information to inform the diagnosis and treatment of clinical patients? First, some links between connectome and consequential brain dynamics might manifest themselves through changes of global network features despite the variability in the changes of individual connections. Examples are global topological changes, observed through diffusion tensor imaging, in remitted geriatric depression and amnestic mild cognitive impairment (Bai et al., 2012). However, the same global changes, say a deviation from the brain's small-world organization towards random or regular connectivity (Reijneveld et al., 2007), could be observed across diseases and therefore limit their use as a classifier for brain diseases.

Second, some changes might be so widespread that they affect the majority or all of the brain regions leaving fewer degrees of freedom for variability in connectomes. The overall pattern of altered structural connectivity in schizophrenia patients (Skudlarski et al., 2010), along with resulting functional connectivity changes (Fornito et al., 2012), would be one example for this case.

Third, changes that are linked to a brain disease might only affect specific circuits in the network. In that way, while the strength of most connections also varies in healthy controls, more consistent changes to specific fiber tracts would be expected for patients. As a consequence, changes in selected circuits would be common for a group of patients but a consistent change for all fiber tracts of a circuit would not occur in control subjects. While this is a potentially powerful approach it does need *a priori* knowledge about the affected circuit. Such circuits might be identified by large-cohort studies in patients or through 'knock-out' studies, e.g. using transcranial magnetic stimulation (Hilgetag et al., 2001), in healthy subjects.

Finally, I would propose a novel approach to deal with the variability in brain disorders, which is the use of computer simulations of brain activity, based on the connectivity in individual patients. Such simulations are already emerging as a way to understand the structural correlates of dynamical changes and disease progression (Cabral et al., 2012; Deco et al., 2011; Raj et al., 2012). As shown above, multiple structural connectivity changes might lead to the same changes in brain dynamics, patient behavior, or clinical test scores. Simulating the activity in the brain of individual patients can inform us about the expected behavioral features and thus about the presence or absence of one sub-type of brain disorder. These models can go beyond the observation of patterns in the recordings of brain activity as simulated dynamics could include more complex models. For example, a model based on structural connectivity might include simulated activity of individual neurons or local circuits, which cannot be observed by non-invasive neuroimaging.

Using simulations in a clinical setting has several benefits. First, simulated behavioral features can be mapped to brain activity in patients that is available through fMRI, PET, MEG, EEG, ECoG, or recordings in resected tissue (Roopun et al., 2010), depending on the disease. Second, the simulated behavior can be compared with the experimentally obtained behavior to validate and constrain a model: simulated activity can be compared with the clinical recordings of a patient. Third, observing dynamics in networks opens up the possibility to use the tools of nonlinear dynamics and time series analysis to find patterns that could be biomarkers for a given disease. Importantly, changes in brain dynamics might be visible even in cases where the structural connectivity is not significantly different from that of a healthy control group. Such simulations are becoming available both at the local (Blue Brain Project, (Markram, 2006)) and global level (Virtual Brain Project, (Jirsa et al., 2010)) and will be support through the Human Brain Project and other initiatives.

In conclusion, there is a large number of underlying structural connectome changes that might lead to the same functional and behavioral changes in healthy subjects and patients. This variety makes the detection of a brain disorder—not just the classification of the type of disorder (Hyman, 2010)—difficult. We propose the use of computer models to use the simulated dynamics (dynamic connectome) based on structural connectivity, rather than the directly measured structural connectivity alone, as a biomarker. In the same way that biology has moved from genes to gene expression data, the use of dynamic connectomes, observing or simulating activity in neural circuits, opens up future potential for clinical applications.

**ACKNOWLEDGEMENTS**
I would like to thank Drs Simon Eickhoff (Düsseldorf) and Stephen Jackson (Nottingham) for inspiring me to work on this question following a discussion at the Fusion Workshop at Korea University. This work was supported by the WCU program through the National Research Foundation of Korea funded by the Ministry of Education, Science and Technology (R32-10142), the CARMEN e-science project (http://www.carmen.org.uk) as well as another project funded by EPSRC (EP/ K026992/1).

**REFERENCES**
Achacoso, T.B., and Yamamoto, W.S. (1992). AY's Neuroanatomy of *C. elegans* for Computation (Boca Raton, FL: CRC Press).
Addington, A.M., and Rapoport, J.L. (2012). Annual research review: impact of advances in genetics in understanding developmental psychopathology. J Child Psychol Psychiatry *53*, 510-518.
Bai, F., Shu, N., Yuan, Y., Shi, Y., Yu, H., Wu, D., et al. (2012). Topologically Convergent and Divergent Structural Connectivity Patterns between Patients with Remitted Geriatric Depression and Amnestic Mild Cognitive Impairment. Journal of Neuroscience *32*, 4307-4318.
Bullmore, E., and Sporns, O. (2009). Complex brain networks: graph theoretical analysis of structural and functional systems. Nat Rev Neurosci *10*, 186-198.
Cabral, J., Hugues, E., Kringelbach, M.L., and Deco, G. (2012). Modeling the outcome of structural disconnection on resting-state functional connectivity. Neuroimage *62*, 1342-1353.
Costa, L.d.F., Rodrigues, F.A., Travieso, G., and Villas Boas, P.R. (2007). Characterization of complex networks: A survey of measurements. Advances in Physics *56*, 167-242.
Deco, G., Jirsa, V.K., and McIntosh, A.R. (2011). Emerging concepts for the dynamical organization of resting-state activity in the brain. Nature Reviews Neuroscience *12*, 43-56.
DeFelipe, J. (2010). From the Connectome to the Synaptome: An Epic Love Story. Science *330*, 1198-1201.

Felleman, D.J., and van Essen, D.C. (1991). Distributed hierarchical processing in the primate cerebral cortex. Cereb. Cortex *1*, 1-47.
Fornito, A., Zalesky, A., Pantelis, C., and Bullmore, E.T. (2012). Schizophrenia, neuroimaging and connectomics. Neuroimage *62*, 2296-2314.
Hilgetag, C.C., and Barbas, H. (2006). Role of Mechanical Factors in the Morphology of the Primate Cerebral Cortex. PLoS Computational Biology *2*, e22.
Hilgetag, C.C., Theoret, H., and Pascual-Leone, A. (2001). Enhanced visual spatial attention ipsilateral to rTMS-induced 'virtual lesions' of human parietal cortex. Nature Neurosci. *4*, 953-957.
Hyman, S.E. (2010). The diagnosis of mental disorders: the problem of reification. Annu Rev Clin Psychol *6*, 155-179.
Jbabdi, S., and Johansen-Berg, H. (2011). Tractography: Where Do We Go from Here? Brain Connectivity *1*, 169-183.
Jirsa, V.K., Sporns, O., Breakspear, M., Deco, G., and McIntosh, A.R. (2010). Towards the virtual brain: network modeling of the intact and the damaged brain. Archives italiennes de biologie *148*, 189-205.
Kaiser, M. (2011). A Tutorial in Connectome Analysis: Topological and Spatial Features of Brain Networks. Neuroimage *57*, 892-907.
Kaiser, M., Hilgetag, C.C., and van Ooyen, A. (2009). A simple rule for axon outgrowth and synaptic competition generates realistic connection lengths and filling fractions. Cereb Cortex *19*, 3001-3010.
Kendler, K.S., Aggen, S.H., Knudsen, G.P., Roysamb, E., Neale, M.C., and Reichborn-Kjennerud, T. (2011). The structure of genetic and environmental risk factors for syndromal and subsyndromal common DSM-IV axis I and all axis II disorders. Am J Psychiatry *168*, 29-39.
Markram, H. (2006). The blue brain project. Nature Rev. Neurosci. *7*, 153-160.
Nisbach, F., and Kaiser, M. (2007). Developmental time windows for spatial growth generate multiple-cluster small-world networks. European Physical Journal B *58*, 185-191.
Price, C.J., and Friston, K.J. (2002). Degeneracy and cognitive anatomy. Trends Cog. Sci. *6*, 416-421.
Raj, A., Kuceyeski, A., and Weiner, M. (2012). A Network Diffusion Model of Disease Progression in Dementia. Neuron *73*, 1204-1215.
Reijneveld, J.C., Ponten, S.C., Berendse, H.W., and Stam, C.J. (2007). The application of graph theoretical analysis to complex networks in the brain. Clin Neurophysiol *118*, 2317-2331.
Roopun, A.K., Simonotto, J.D., Pierce, M.L., Jenkins, A., Nicholson, C., Schofield, I.S., et al. (2010). A nonsynaptic mechanism underlying interictal discharges in human epileptic neocortex. Proc Natl Acad Sci U S A *107*, 338-343.
Skudlarski, P., Jagannathan, K., Anderson, K., Stevens, M.C., Calhoun, V.D., Skudlarska, B.A., et al. (2010). Brain Connectivity Is Not Only Lower but Different in Schizophrenia: A Combined Anatomical and Functional Approach. Biological Psychiatry *68*, 61-69.
Sporns, O., Chialvo, D.R., Kaiser, M., and Hilgetag, C.C. (2004). Organization, development and function of complex brain networks. Trends Cogn Sci *8*, 418-425.
Tononi, G., Sporns, O., and Edelman, G.M. (1999). Measures of degeneracy and redundancy in biological networks. Proc Natl Acad Sci U S A *96*, 3257-3262.
Turing, A.M. (1952). The Chemical Basis of Morphogenesis. Philosophical Transactions of the Royal Society of London. Series B, Biological Sciences *237*, 37-72.
van den Heuvel, M.P., Kahn, R.S., Goni, J., and Sporns, O. (2012). High-cost, high-capacity backbone for global brain communication. Proc Natl Acad Sci U S A *109*, 11372-11377.
Van Essen, D.C. (1997). A tension-based theory of morphogenesis and compact wiring in the central nervous system. Nature *385*, 313-318.
White, J.G., Southgate, E., Thomson, J.N., and Brenner, S. (1986). The structure of the nervous system of the nematode *Caenorhabditis elegans*. Phil. Trans. R. Soc. of London B *314*, 1.
Young, M.P. (1992). Objective Analysis of the Topological Organization of the Primate Cortical Visual System. Nature *358*, 152-155.

**FIGURES**

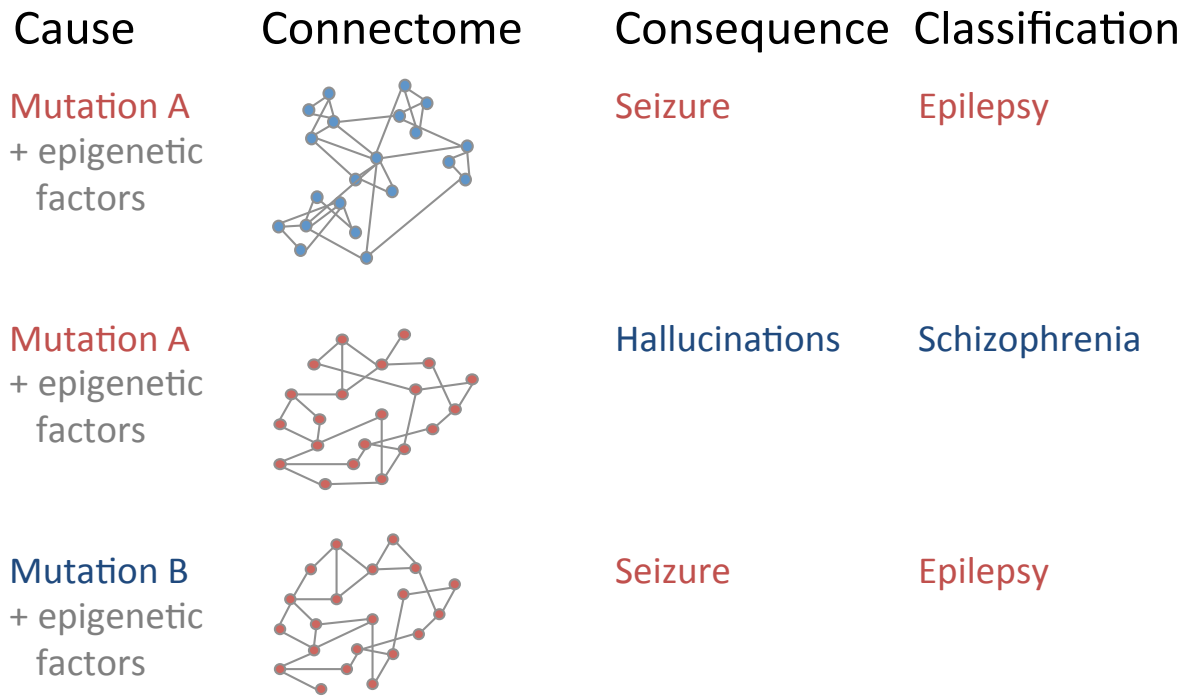

| Cause | Connectome | Consequence | Classification |
|---|---|---|---|
| Mutation A + epigenetic factors | | Seizure | Epilepsy |
| Mutation A + epigenetic factors | | Hallucinations | Schizophrenia |
| Mutation B + epigenetic factors | | Seizure | Epilepsy |

**Figure 1.** Mapping between underlying developmental causes of connectome changes, ranging from genetic factors to spatiotemporal epigenetic factors, to resulting brain connectivity ('connectome'), observable network behavior ('consequence'), and final disease classification. Similar patterns within each of the four categories are shown in red. Note that both genetic patterns and network features alone may be insufficient to inform the clinical diagnosis of a disease: First, the same genetic mutation A might lead to a different connectivity due to different epigenetic factors. Second, different genetic mutations A & B could lead to the same connectivity due to additional factors. Third, the same connectivity might lead to different behavior and disease classification due to changes that solely affect the anatomical organization within individual nodes.